\documentclass[10pt,conference]{IEEEtran}
\IEEEoverridecommandlockouts
\usepackage{colortbl}
\usepackage{enumitem}
\usepackage{bbding}
\usepackage{amsmath}
\usepackage{amsfonts}
\usepackage{listings}
\usepackage{hyperref}
\usepackage{multirow}
\usepackage{tcolorbox}
\usepackage{listings}
\usepackage{cprotect}
\usepackage{hyperref}
\usepackage{tikz}

\usepackage{circledtext}
\usepackage{float}
\usepackage{pgfplots}
\pgfplotsset{compat=1.18}
\usepackage{subcaption}

\lstdefinestyle{python}{
    language=Python,
    frame=lines,
    basicstyle=\ttfamily\scriptsize,
    keywordstyle=\color{blue},
    commentstyle=\color{teal},
    stringstyle=\color{red},
    numbers=left,
    numberstyle=\tiny\color{gray},
    stepnumber=1,
    numbersep=5pt,
    showstringspaces=false,
    breaklines=true,
    breakatwhitespace=true,
    tabsize=4,
    captionpos=b
}
\newcommand{\toolname}{\texttt{RepoMasterEval}~}

\def\BibTeX{{\rm B\kern-.05em{\sc i\kern-.025em b}\kern-.08em
    T\kern-.1667em\lower.7ex\hbox{E}\kern-.125emX}}

\begin{document}

%
\title{RepoMasterEval: Evaluating Code Completion via Real-World Repositories}

%

\makeatletter

\newcommand{\linebreakand}{%
  \end{@IEEEauthorhalign}
  \hfill\mbox{}\par
  \mbox{}\hfill\begin{@IEEEauthorhalign}
}
\makeatother

\author{
\IEEEauthorblockN{Qinyun Wu\IEEEauthorrefmark{1}\textsuperscript{$\clubsuit$}, Chao Peng\IEEEauthorrefmark{1}\textsuperscript{$\heartsuit$}\textsuperscript{$\clubsuit$}, Pengfei Gao\IEEEauthorrefmark{1}, Ruida Hu\IEEEauthorrefmark{2}\textsuperscript{$\spadesuit$}, Haoyu Gan\IEEEauthorrefmark{1}, Bo Jiang\IEEEauthorrefmark{1}, Jinhe Tang\IEEEauthorrefmark{1}\textsuperscript{$\spadesuit$}, Zhiwen Deng\IEEEauthorrefmark{1},\\Zhanming Guan\IEEEauthorrefmark{1}, Cuiyun Gao\IEEEauthorrefmark{2}\textsuperscript{$\heartsuit$}, Xia Liu\IEEEauthorrefmark{1}, Ping Yang\IEEEauthorrefmark{1}}
\IEEEauthorblockA{
\IEEEauthorrefmark{1} ByteDance, China \\
\IEEEauthorrefmark{2} Harbin Institute of Technology, Shenzhen, China \\
\{wuqinyun, pengchao.x, gaopengfei.se, ganhaoyu, jiangbo.jacob, tangjinhe, dengzhiwen.11,\\linlandong, yangping.cser\}@bytedance.com \\
200111107@stu.hit.edu.cn, gaocuiyun@hit.edu.cn
}
}

\maketitle

\begingroup
\renewcommand\thefootnote{$\clubsuit$}
\footnotetext{Equal contribution.}
\renewcommand\thefootnote{$\heartsuit$}
\footnotetext{Corresponding authors.}
\renewcommand\thefootnote{$\spadesuit$}
\footnotetext{Work down during an internship at ByteDance.}
\endgroup

\begin{abstract}
    With the growing reliance on automated code completion tools in software development, the need for comprehensive
evaluation benchmarks has become critical.
Existing benchmarks focus more on code completion
in function and class level by providing
text descriptions to prompt the model.
By contrast, such descriptive prompt is commonly unavailable in real development and code completion can occur in wider range of situations such as in the middle of a function or a code block.
These limitations makes existing evaluation benchmarks poorly align with the practical scenarios of code completion tools.
In this paper, we propose \toolname, a novel benchmark for evaluating code completion models constructed from real-world repositories.
Each benchmark datum is generated by masking a code snippet (ground truth) from one source code file with existing test suites.
To improve test accuracy of model generated code, we employ mutation testing to measure the effectiveness of the test cases and we manually crafted new test cases for those test suites with low mutation score.
Our empirical evaluation on 10 state-of-the-art models shows that test argumentation is critical in improving the accuracy of the benchmark and \toolname is able to report variance in model performance in real-world scenarios.
The deployment of \toolname also revealed that the benchmark is useful to give accurate feedback during model training and the score is in high correlation with the model's performance in practice.
\end{abstract}


\section{Introduction}
\label{sec:intro}
The rapid advancements in large language models (LLMs), prompt strategies, context retrieval algorithms, and tool availability have significantly improved the capability and popularity of automated code completion and generation tools, such as GitHub Copilot~\cite{copilot}, MarsCode~\cite{marscode} and Codeium~\cite{codeium}.
These tools, often integrated as plugins of Integrated Development Environments (IDEs), predict code snippets at the cursor's location based on existing code, comments, and repository context, with users accepting the suggestions by pressing the \emph{Tab} key.  
Prominent LLMs, such as GPT family models~\cite{achiam2023gpt}, DeepSeek Coder~\cite{guo2024deepseek}, and StarCoder~\cite{lozhkov2024starcoder}, have demonstrated impressive capabilities by leveraging massive corpora and instructions during training.

To comprehensively evaluate the performance of these emerging LLMs, several benchmarks have been proposed, such as HumanEval~\cite{chen2021evaluating} and MBPP~\cite{austin2021program}.
These benchmarks typically consist of prompts with function signatures, expected behavior, and example inputs/outputs for the model to generate code snippets, alongside evaluation metrics to determine the correctness of the generated code. 
For instance, HumanEval~\cite{chen2021evaluating} and ClassEval~\cite{du2023classeval} use test cases and calculate the pass rate of models' generated code, while RepoBench~\cite{liu2023repobench} compares the generated code with the ground truth based on code similarity metrics~\cite{svyatkovskiy2020intellicode}. 

Although these benchmarks have been instrumental in understanding and comparing the performance of different models in code generation tasks, they are not suitable for real-world code completion due to the following limitations:

\textbf{\textit{Challenge 1: Limited Scenarios.}}  Primarily, current code completion benchmarks \cite{chen2021evaluating,liu2023repobench,li2024evocodebench,du2023classeval} focus on relatively simple or straightforward scenarios, such as statement-level, function-level, or class-level completion. These scenarios typically involve completing a single code unit (e.g., a statement, function, or class) in isolation. However, in real-world software development, code completion tasks can occur at any point, including in the middle of code blocks, with or without subsequent code, and are not limited to complete syntactical or semantic units in isolation.

Additionally, there is a lack of analysis on covered scenarios for existing benchmarks.
Increasing the number of repositories in the evaluation set can introduce biases. Many tasks from different repositories involve basic utility functions. If the evaluation focuses on these repetitive tasks, it mainly tests the same abilities and may miss the model's weaknesses. This can lead to a gap between evaluation results and real-world performance, emphasizing the need for a benchmark that better reflects real-world scenarios.


\textbf{\textit{Challenge 2: Limited Test Suite Quality Assurance.}} 
To fairly evaluate model performance on masked code, it is essential to utilize unit tests to execute the predicted code and verify its accuracy. This approach is necessary because code snippets that achieve the same functionality can vary significantly and exhibit low similarity. Currently, very few benchmarks use unit test pass rates as the metric. Among the limited benchmarks that do incorporate unit tests, test cases are often inadequate, sometimes passing even when the model’s predictions are incorrect.


To address these limitations, we introduce \toolname, a novel benchmark designed to evaluate code completion models in more realistic and challenging scenarios. Our benchmark is constructed from real-world Python and TypeScript repositories on GitHub and includes an analysis of the categories of tasks covered, ensuring a fair distribution across those categories.

Each data point in the benchmark involves masking a code snippet from a source file and prompting the model to predict the masked snippet using both prefix and suffix contexts, as well as repository-level context retrieved based on the BM25~\cite{ding2024crosscodeeval} algorithm. By incorporating a diverse range of code completion scenarios, including those that occur in the middle of code blocks, \toolname aims to offer a more accurate and practical assessment of the capabilities of language models. This setup simulates the Fill-In-the-Middle task, which is more representative of real-world production.

To ensure the accuracy and robustness of the benchmark, we employed mutation testing and manual test case crafting. Mutation testing generates mutants from the ground truth code snippets and executing the test cases to calculate the mutation score. For code snippets with low mutation scores, we manually crafted additional test cases to enhance test effectiveness. This process ensures that the generated code snippets are indeed correct and robust.


Our comprehensive evaluation of various state-of-the-art models on \toolname revealed significant insights into the performance and applicability of these models in real-world code completion tasks. We found that models generally performed better on simpler, more isolated tasks of benchmarks like HumanEval but faced challenges with the more complex, context-rich scenarios presented by \toolname. For instance, while GPT-4 achieved high scores on HumanEval, its performance dropped significantly on \toolname, highlighting the benchmark’s ability to reflect real-world complexities.

Moreover, \toolname provides granular insights based on model performance across different data points. For example, programming basics typically represent an area where most models score the best, indicating a proficiency in foundational coding tasks. This performance analysis helps identify specific strengths and weaknesses of the models, offering a clearer picture of their practical applicability in diverse coding environments.

Additionally, the effectiveness of manual test augmentation was evident, with substantial improvements in mutation scores for both programming languages and decrease in models' pass rate, though code coverage remained largely unchanged. 
This indicates that augmented tests enhance robustness, posing greater challenges for the models. Furthermore, a positive correlation was observed between the model’s performance on \toolname and its online acceptance rate during a one-month deployment period, validating the benchmark’s relevance and effectiveness in practical applications.

Our findings underscore the need of \toolname to accurately assess and guide the optimization of LLMs for real-world software development tasks and we call for the software engineering research community to build more practical and complicated code completion benchmarks.

In summary, this paper makes the following contributions:

\begin{itemize}[leftmargin=*]
    \item We introduce a novel benchmark for evaluating code completion models in real and complex scenarios with practical settings.
    \item We employ mutation testing and manual test case crafting to ensure the accuracy and robustness of the benchmark.
    \item We conduct the first industrial study and demonstrate the deployment of \toolname and the correlation between model performance in benchmarking and practice.
\end{itemize}

\section{Background and Motivation}
\label{sec:background}
In this section, we briefly introduce code completion, LLMs for these tasks and motivate our work by revisiting existing benchmarks.

\begin{table*}[!htbp]
\caption{Comparison between Existing Benchmarks and \toolname}
\centering
\small
\begin{tabular}{cccccccc}
\hline
Benchmark & Year & \begin{tabular}[c]{@{}c@{}}Human   \\ Annotation\end{tabular} & \begin{tabular}[c]{@{}c@{}}Real  \\ Repos\end{tabular} & \begin{tabular}[c]{@{}c@{}}Diversity   \\ Analysis\end{tabular} & Prompt & Granularity & Metrics \\ \hline
HumanEval & 2021 & \CheckmarkBold & \XSolidBrush & N/A & DS & Function & Testing \\ 
AixBench & 2023 & \XSolidBrush & \XSolidBrush & N/A & TR \& FS & Function & Testing \\ 
RepoBench & 2023 & \XSolidBrush & \CheckmarkBold & N/A & Code \& RC & Line & Similarity \\ 
RepoEval & 2023 & \XSolidBrush & \CheckmarkBold & N/A & Code \& RC & Line \& Func. Body & Similarity \& Testing \\
ClassEval & 2023 & \CheckmarkBold & \XSolidBrush & N/A & DS \& CS & Class & Testing \\
CoderEval & 2023 & \XSolidBrush & \CheckmarkBold & N/A & DS & Function & Testing \\
EvoCodeBench & 2023 & \CheckmarkBold & \CheckmarkBold & Repo   level & DS \& CF & Function & Testing \\ 
CrossCodeEval & 2023 & \CheckmarkBold & \CheckmarkBold & N/A & Code \& RC & Line & Similarity \\ 
\textbf{RepoMasterEval} & 2024 & \CheckmarkBold & \CheckmarkBold & Code   snippet   level & Code \& RC & Mixed & Enhanced   Testing \\ \hline
\end{tabular} \\
DS: doc string, TR: text requirements, FS: function signature, RC: repository context, CS: class skeleton, CF: current file
\label{table:existing-benchmarks}
\end{table*}

\subsection{Large Language Models for Code}

LLMs leverage massive datasets and sophisticated training techniques to produce coherent and contextually relevant code snippets.
Recent advances in LLMs, especially those trained on code, have revolutionized the field of automated software engineering, providing significant enhancements in productivity and accuracy.

\textbf{General LLMs} have shown exceptional performance in various development tasks. For instance, GPT-4 achieves high pass rates on benchmarks like HumanEval. With their extensive training on diverse datasets, these models can generate code snippets and even provide debugging assistance based on natural language prompts.

\textbf{Code-specific LLMs} are trained primarily on code-specific data and often outperform general LLMs in code-related tasks.  Notable examples include Codex~\cite{chen2021evaluating}, DeepSeek Coder~\cite{guo2024deepseek} and StarCoder~\cite{lozhkov2024starcoder}, which have been fine-tuned to generate accurate and contextually appropriate code. These models employ various training objectives, such as next-token prediction or filling-in-the-middle (FIM) techniques.

\subsection{Existing Benchmarks for Code Completion}

To evaluate the performance of these advanced LLMs in code completion, several benchmarks have been developed.
The task is consisted of a natural language description as the input (prompt), and the corresponding code acts as the ground truth output (canonical solution).
In terms of metrics, exact match and code similarity methods compare the model output with the ground truth while passing rate (Pass@k) executes the model output against test cases to assess the correctness of the generated code.

We revisit existing code completion benchmarks that are used actively according to studies conducted by~\cite{du2023classeval, li2024evocodebench}. As summarized in Table~\ref{table:existing-benchmarks}, these benchmarks fall short in assessing practical code completion scenarios encountered in real-world software development due to the following reasons:

\begin{enumerate}[leftmargin=*]
\item \textbf{Focus on Single Code Units:} These benchmarks evaluate the completion of isolated code units, such as individual functions or classes, rather than more diverse, interdependent code structures such as loop body, part of the function, etc. This approach limits the assessment to simpler tasks, which may not fully exploit the capabilities of modern LLMs capable of handling longer sequences and more intricate dependencies.
\item \textbf{Limited and Impractical Scenarios:} Existing benchmarks typically cover simple scenarios or lack diversity analysis, leading to biased evaluation results and making it easier for models to be fine-tuned to boost metrics artificially. Additionally, these benchmarks often provide impractical contextual information. Prompts in the form of documentation strings and text descriptions focus on the functionality of the code to be implemented but lack additional context from surrounding code and relevant source files. In real-world scenarios, code completion tasks often 
occur in broader and more complex codebases where functions and methods are interdependent.

\item \textbf{Test Suite Quality:} The reliance on predefined test cases for evaluation can result in insufficient assessment of robustness, as these test cases might not cover all possible edge cases. HumannEval+~\cite{liu2024your} examined the test effectiveness of HumanEval via mutation testing and revealed the ineffectiveness of existing test cases of HumanEval.
\item \textbf{Lack of Empirical Correlation Studies:} There is a gap in research examining the correlation between benchmark performance and real-world usability, making it difficult to determine the practical effectiveness of these benchmarks.
\end{enumerate}

\subsection{Motivation for a Code Completion Benchmark}

As existing benchmarks are inadequate for evaluating more practical code completion tasks, such as generating longer and compound code units consisting of multiple interdependent methods, we propose \toolname designed specifically for code completion tasks, to cover more realistic and challenging scenarios. 
\toolname incorporates rich contextual information from real-world repositories and employs mutation testing and manual test case crafting to ensure accuracy and robustness.
This approach offers a more comprehensive and practical framework for assessing LLM performance in real-world software development.


\section{Approach}
\label{sec:approach}
In this section, we present the overview of \toolname in Section~\ref{sec:benchmark-overview}) and present data collection (Section~\ref{sec:data-source}), task construction and test suite augmentation via mutation testing  (Section~\ref{sec:task-construction}) and evaluation process (Section~\ref{sec:evaluation-metrics}) using this example. We also discuss the diversity of the benchmark in Section~\ref{sec:diversity-study}.

\subsection{Benchmark Overview}
\label{sec:benchmark-overview}

\toolname is designed to provide a comprehensive and realistic evaluation of code completion models, reflecting the complex and varied scenarios encountered in real-world software development. As summarized in Table~\ref{table:benchmark-overview}, each coding task consists of the following key components:

\circledtextset{resize=real}
\begin{itemize}[leftmargin=*]
    \item[] \circledtext*[height=2ex,charshrink=0.8]{1} \textbf{Prefix:} Code that appears before the masked snippet. It provides essential context for the code completion task, helping the model understand the surrounding code environment.
    \item[] \circledtext*[height=2ex,charshrink=0.8]{2} \textbf{Masked Code:} The masked code snippet that the model needs to generate. This serves as the correct output that models are evaluated against.
    \item[] \circledtext*[height=2ex,charshrink=0.8]{3} \textbf{Suffix:} The code that follows the masked snippet. This additional context is crucial for models to generate accurate and contextually appropriate code completions.
    \item[] \circledtext*[height=2ex,charshrink=0.8]{4} \textbf{Retrieved Information:} Contextual information retrieved from the repository using the BM25 algorithm. This includes relevant code snippets, comments, and documentation that can help the model make better predictions.
    \item[] \circledtext*[height=2ex,charshrink=0.8]{5} \textbf{Test Cases:} A set of test cases used to evaluate the functional correctness of the generated code. These test cases are designed to cover various edge cases and ensure that the generated code is robust and performs as expected.
\end{itemize}

The model is prompted with \circledtext*[height=2ex,charshrink=0.8]{1} \textbf{Prefix}, \circledtext*[height=2ex,charshrink=0.8]{3} \textbf{Suffix} and \circledtext*[height=2ex,charshrink=0.8]{4} \textbf{Retrieved Information} to generate the \circledtext*[height=2ex,charshrink=0.8]{2} masked code, which is placed to the original location and executed against \circledtext*[height=2ex,charshrink=0.8]{5} test cases. 

This structure simulates real-world code completion in the IDE and addresses the need for a more realistic and comprehensive evaluation of code completion models.
Unlike traditional benchmarks that focus on generating isolated code units with descriptive text prompt,
\toolname emphasizes generating code snippets within a broader context, simulating real-world development.



\newsavebox{\lstboxprefix}
\begin{lrbox}{\lstboxprefix}
\begin{minipage}[t]{0.5\textwidth}
\begin{lstlisting}[language=Python,frame=no,numbers=none]
# 14 lines of code ommited
def get_or_create_composite_action(path: str) -> "CompositeAction":
    """Used when need to create relations with another action.
    If action wasn't indexed yet, we create a stub node,
    that will be enriched eventually.
    """
    ca = CompositeAction(None, path)
    obj = Config.graph.get_object(ca)
    if not obj:
        # This is a legitimate behavior.
        # Once the action will be indexed, the node will be enriched.
\end{lstlisting}
\end{minipage}
\end{lrbox}

\newsavebox{\lstboxsuffix}
\begin{lrbox}{\lstboxsuffix}
\begin{minipage}[t]{0.5\textwidth}
\begin{lstlisting}[language=Python,frame=no,numbers=none]
class CompositeActionInput(GraphObject):
    __primarykey__ = "_id"

    _id = Property()
    name = Property()
    default = Property()
    description = Property()
    required = Property()
    url = Property()
    path = Property()

    def __init__(self, _id: str, path: str):
        self._id = _id
        self.path = path
# 128 lines of code ommited
\end{lstlisting}
\end{minipage}
\end{lrbox}

\newsavebox{\lstboxgt}
\begin{lrbox}{\lstboxgt}
\begin{minipage}[t]{0.5\textwidth}
\begin{lstlisting}[language=Python,frame=no,numbers=none]
    Config.graph.push_object(ca)
    obj = ca
return obj
\end{lstlisting}
\end{minipage}
\end{lrbox}

\newsavebox{\lstboxretreived}
\begin{lrbox}{\lstboxretreived}
\begin{minipage}[t]{0.5\textwidth}
\begin{lstlisting}[language=Python,frame=no,numbers=none]
# /src/workflow_components/workflow.py

# 58 lines of code ommited

# /src/workflow_components/workflow.py

# 19 lines of code ommited

def get_or_create_workflow(path: str) -> "Workflow":
    """Used when need to create relations with another workflow.
    If workflow wasn't indexed yet, we create a stub node,
    that will be enriched eventually.
    """
    w = Workflow(None, path)
    obj = Config.graph.get_object(w)
    if not obj:
        # This is a legitimate behavior.
        # Once the workflow will be indexed, the node will be enriched.
        Config.graph.push_object(w)
        obj = w
    return obj


# 14 lines of code ommited
\end{lstlisting}
\end{minipage}
\end{lrbox}

\newsavebox{\lstboxut}
\begin{lrbox}{\lstboxut}
\begin{minipage}[t]{0.5\textwidth}
\begin{lstlisting}[language=Python,frame=no,numbers=none]
# import statements omitted for brevity

load_test_config()

def test_get_or_create_composite_action():
    # Arrange
    path = "test_path"
    ca = composite_action.CompositeAction(
            None, 
            path)
    Config.graph.get_object = MagicMock(return_value=None)
    Config.graph.push_object = MagicMock()

    # Act
    result = composite_action.
      get_or_create_composite_action(path)

    # Assert
    assert result == ca
    Config.graph.get_object.
      assert_called_once_with(ca)
    Config.graph.push_object.
      assert_called_once_with(ca)
\end{lstlisting}
\end{minipage}
\end{lrbox}

\circledtextset{resize=real}
\begin{table*}[!htbp]
    \caption{\toolname Overview}
    \label{table:benchmark-overview}
    \footnotesize{
        \begin{tabular}{cc}
        \multicolumn{2}{c}{
            \begin{tabular}[c]{@{}c@{}}\normalsize{\textbf{Task}: With \circledtext*[height=2ex,charshrink=0.8]{1}prefix, \circledtext*[height=2ex,charshrink=0.8]{3}suffix and \circledtext*[height=2ex,charshrink=0.8]{4}retrieved information, generate code snippet (\circledtext*[height=2ex,charshrink=0.8]{2})} \\ \normalsize{\textbf{Metrics:} Execute model generated code against \circledtext*[height=2ex,charshrink=0.8]{5}test cases}\end{tabular}
        }                            \\ \hline
        \multicolumn{1}{|l|}{
            \begin{tabular}[c]{@{}l@{}}\textbf{Metadata}\\ \textbf{Language:} Python\\ \textbf{Related Domain:} Multimedia - Image Processing\end{tabular}
        } &
        \multicolumn{1}{l|}{
            \begin{tabular}[c]{@{}l@{}}\circledtext*[height=2ex,charshrink=0.8]{2} \textbf{Masked Code (To be completed by the model)}\\ \usebox{\lstboxgt}\end{tabular}
        } \\ \hline
        \multicolumn{1}{|l|}{
            \begin{tabular}[c]{@{}l@{}}\circledtext*[height=2ex,charshrink=0.8]{1} \textbf{Prefix}\\ \usebox{\lstboxprefix}\end{tabular}
        } & \multicolumn{1}{l|}{
            \begin{tabular}[c]{@{}l@{}}\circledtext*[height=2ex,charshrink=0.8]{3} \textbf{Suffix}\\ \usebox{\lstboxsuffix}\end{tabular}
        } \\ \hline
        \multicolumn{1}{|l|}{
            \begin{tabular}[c]{@{}l@{}}\circledtext*[height=2ex,charshrink=0.8]{4} \textbf{Retrieved Information}\\ \usebox{\lstboxretreived}\end{tabular}
        } & \multicolumn{1}{l|}{
            \begin{tabular}[c]{@{}l@{}}\circledtext*[height=2ex,charshrink=0.8]{5} \textbf{Test Cases}\\ \usebox{\lstboxut}\end{tabular}
        } \\ \hline
        \end{tabular}
    }
    \end{table*}

\subsection{Data Source}
\label{sec:data-source}

To achieve the commitment to realism, we select active and continuously updated GitHub repositories as the data source. To mitigate the potential for data leakage, the benchmark exclusively incorporates repositories inaugurated post-March 2023. While data leakage is a challenging issue to fully address, our approach largely automates task construction. This makes it straightforward to update tasks using recently created repositories, thereby further reducing the risk of data leakage.

To align with a strict quality standard, \toolname employs rigorous filtering criteria: \circledtext[height=2ex,charshrink=0.8]{1} Each repository must have gained at least 100 stars, ensuring a baseline level of community endorsement and visibility. \circledtext[height=2ex,charshrink=0.8]{2} Recognizing the critical role of unit test pass rates in the evaluation metrics, only repositories with a proven track record of successful unit test executions are included. This is confirmed through the presence of test files, automated test execution pipelines, and our additional manual verification that all tests pass.

Repositories shown in Table~\ref{table:repositories} are selected under this criteria and human inspection.

\begin{table}[!thbp]
    \caption{Repositories included in \toolname}
    \label{table:repositories}
    \small
    \begin{tabular}{cccc}
    \hline
    Language                    & Repository Name                                   & Start Date & \# Stars \\ \hline
    \multirow{3}{*}{Python}     & \href{https://github.com/gpt-engineer-org/gpt-engineer}{gpt-engineer} & 2023-04-29   & 48.7k       \\
                                & \href{https://github.com/aurelio-labs/semantic-router}{semantic-router}  & 2023-10-30   & 613         \\
                                & \href{https://github.com/CycodeLabs/raven}{raven}              & 2023-09-12   & 474        \\ \hline
    \multirow{3}{*}{TypeScript} & \href{https://github.com/epicweb-dev/epic-stack}{epic-stack}        & 2023-05-04      &  3.7k     \\
                                & \href{https://github.com/lobehub/lobe-chat}{lobe-chat}            & 2023-05-21      &21.1k     \\
                                & \href{https://github.com/ant-design/ant-design-web3}{ant-design-web3}    & 2023-08-18   & 598         \\ \hline
    \end{tabular}
\end{table}

\subsection{Task Construction}
\label{sec:task-construction}

For the evaluation to be effective, it is crucial that the ``hole'' (masked code) created for the task is 
covered by the original test set. This coverage ensures that the model’s completion can be accurately tested through existing tests.

The process first executes original tests within the repository to establish a baseline using the original code content. Subsequently, a coverage report is generated that documents which lines of code are covered by the test suite. Analyzing this report allows for the precise identification of code segments covered by tests. 

Listing~\ref{lst:coverage_json} shows an example testing report for the \texttt{Raven} repository.
The \texttt{utils.py} module was executed with 28 lines covered, as specified by the respective line numbers. Notably, Lines 14, 15, and 16 form the \texttt{get\_dependencies\_in\_code} function which spans these lines (illustrated in Listing~\ref{lst:code_sample}).
This suggests that at least one of the existing test cases encompasses this function, evaluating its functionality either directly or indirectly. Consequently, this can be identified as a target ``hole'' for \toolname.

The tasks are designed to cover single line completion, block-level completion, and function-level completion, ensuring comprehensive coverage of real-world development scenarios. For Python, the distribution of tasks is as follows: 69.5\% are at the block level, 16.5\% are at the line level, and 13\% are at the function level. In addition, TypeScript tasks are distributed with 45\% at the block level, 11.6\% at the line level, and 43.4\% at the function level. When constructing function-level tasks, we include method names. We have confirmed that the retrieved context and suffix provide hints for models to predict method names, simulating the real-world implementation where the function is called in the retrieved context.

\begin{lstlisting}[language=Python, label={lst:coverage_json}, caption={Example of a Coverage Report}]
{
  "src/common/utils.py": {
    "executed_lines": [1, 2, 3, 5, 6, 8, 9, 10, 11, 14, 15, 16, 19, 20, 21, 23, 26, 30, 46, 87, 95, 98, 103, 104, 107, 122, 126, 127], 
    "summary": {
      "covered_lines": 28, 
      "num_statements": 59, 
      "percent_covered": 47.45762711864407, 
      "percent_covered_display": "47", 
      "missing_lines": 31, 
      "excluded_lines": 0
    }, 
    "missing_lines": [27, 37, 38, 39, 41, 42, 43, 52, 53, 54, 56, 57, 62, 64, 65, 66, 67, 68, 69, 70, 73, 76, 77, 79, 81, 83, 84, 116, 117, 119, 123], 
    "excluded_lines": []
  }
}
\end{lstlisting}

\begin{lstlisting}[label={lst:code_sample}, caption={Example of a Target Hole (Highlighed with @)}, emph={14-16},language=Python]
...
from src.storage.redis_connection import RedisConnection
from src.config.config import Config
import src.logger.log as log
from urllib.parse import urlparse, parse_qs


@def get_dependencies_in_code(code: str) -> List[str]:@
@    re_fmt = r"\$\{\{\s*([a-zA-Z0-9\-\._]*)\s*\}\}"@
@    return [match.group(1) for match in re.finditer(re_fmt, code)]@
...
\end{lstlisting}

To evaluate and demonstrate the comprehensiveness and diversity of \toolname, it is essential to categorize all target holes based on the functionality they perform within the code snippets or the repository.
This categorization is achieved through labeling by our experienced developers using 6 man-days.
Balancing the need for diversity with the efficiency of task execution, \toolname comprises a total of 288 tasks, (115 Python tasks and 173 TypeScript tasks).
These tasks span across 13 categories, covering a wide range of domains including front-end development, databases, machine learning, and more.
Detailed explanations of these categories and their respective tasks is discussed in Section~\ref{sec:diversity-study}.

After confirming the target holes, it is important to construct a proper prompt for evaluation. To align with the code completion task, the current file's prefix and suffix are essential. Additionally, as \toolname serves as a repository-level evaluation benchmark, it is necessary to incorporate context across files within the current repository. We reuse BM25 repository-level context retrieval algorithm proposed by ~\cite{ding2024crosscodeeval} to retrieve similar code snippets from other files. The current prefix, constrained by the sliding window size, acts as the target code snippet for retrieving other relevant code snippets. These additional snippets are gathered by applying a sliding window across all other files in the repository. The snippets are then ranked based on their similarity scores, and up to five of the highest-ranking snippets are selected to be added to the final prompt. This approach ensures a comprehensive evaluation of model performance when using \toolname .

\subsection{Test Augmentation via Mutation Testing}
\label{sec:mutation-testing}

Mutation testing is a method used to evaluate the effectiveness of test suites by generating defective versions of code, known as mutants, and assessing whether the test cases can detect (“kill”) these mutants. Our test augmentation process involves the following steps:

\begin{enumerate}[leftmargin=*]
    \item \textbf{Generating Defective Versions (Mutants).} We generate various defective versions of the code snippets. These mutants are systematically altered versions of the original code, where specific changes (mutations) are introduced to create potential faults. The goal is to simulate common programming errors and assess whether the existing test cases can identify these faults. Mutation types used in our work is summarized in Table~\ref{table:mutations}.
    \item \textbf{Running Test Suites on Mutants.}  The generated mutants are then subjected to the existing test suites. Each test case in the suite is executed against the mutants to determine whether it can detect the introduced defects. A test case is considered to have “killed” a mutant if it fails when executed on the mutant.
    \item \textbf{Augmenting Test Suites.}  For mutants that are not detected (i.e., not killed) by the current test suites, additional test cases are crafted  using a combination of automated methods (such as GPT-based test generation) and manual annotation by developers. We use a human-LLM collaboration mode, where LLMs are prompted to enhance the test suites based on the the mutation information, and humans fix the generated tests that cannot be compiled. The aim is to enhance the test suite’s ability to detect faults by adding more comprehensive and targeted test cases.
    \item \textbf{Iterative Process} Steps 2 and 3 are repeated iteratively. Each iteration involves running the augmented test suite on the remaining undetected mutants and adding new test cases for any mutants that still survive. This process continues until all mutants are effectively killed by the test suite, indicating a robust and comprehensive set of tests.
\end{enumerate}

\begin{table}[!htbp]
\caption{Mutation Types}
\label{table:mutations}
\scriptsize{
\begin{tabular}{ccc}
\hline
Type & Description & Applicable Languages \\ \hline
AOR                                                     & Arithmetic operator replacement                                                         & Python, TypeScript                                             \\
ASR                                                     & Assignment operator replacement                                                          & Python, TypeScript                                             \\
BCR                                                     & Break/Continue replacement                                                               & Python, TypeScript                                             \\
BOD                                                     & Binary operator deletion                                                                 & Python, TypeScript                                             \\
COD                                                     & Conditional operator deletion                                                            & Python, TypeScript                                             \\
COI                                                     & Conditional operator insertion                                                           & Python, TypeScript                                             \\
CRP                                                     & Constant replacement                                                                     & Python, TypeScript                                             \\
DDL                                                     & Decorator Deletion                                                                       & Python, TypeScript                                             \\
EHD                                                     & Exception handler deletion                                                               & Python, TypeScript                                             \\
IDE                                                     & \begin{tabular}[c]{@{}c@{}}Incremental and decremental\\ operator exchange\end{tabular} & TypeScript                                                     \\
LCR                                                     & Logical connector replacement                                                            & Python, TypeScript                                             \\
LOR                                                     & Logical operator replacement                                                             & Python, TypeScript                                             \\
LSR                                                     & Logical assignment operator replacement                                                  & Python, TypeScript                                             \\
OIL                                                     & One iteration loop                                                                      & TypeScript                                                     \\
ROR                                                     & Relational operator replacement                                                          & Python, TypeScript                                             \\
SDL                                                     & Statement deletion                                                                       & Python, TypeScript                                             \\ \hline
\end{tabular}
}
\end{table}

Specifically, there were 1,105 tests in total, with a mutation score of 58\% for TypeScript projects and 73.6\% for Python projects. After adding 186 tests, the mutation scores improved to 68.2\% and 84.5\%, respectively, as detailed in Table \ref{table:Mutation_Scores}. By employing mutation testing in this manner, we significantly enhance the robustness and adequacy of our test suites. This iterative approach ensures that the test cases are capable of identifying a wide range of potential defects, ultimately leading to more reliable and fault-tolerant code. This methodology not only improves the accuracy of our benchmarks but also provides valuable insights into the areas where the test suites need strengthening, thereby guiding the development of more effective benchmarks.

\begin{table*}[]
\caption{Mutation Scores}
\label{table:Mutation_Scores}
\footnotesize\begin{tabular}{ccccccc}
\hline

Language & Repository &  Existing Tests &  Tests after Augmentation & \% Improvement & Original Mutation Score & New Mutation Score \\ \hline

TypeScript & epic-stack & 16 & 48 & 200.0\% & 44.1\% & 61.9\% \\
TypeScript & ant-design-web3 & 239 & 252 & 5.4\% & 62.1\% & 66.3\% \\
TypeScript & lobe-chat & 608 & 695 & 14.3\% & 61.5\% & 70.2\% \\
Python & gpt-engineer & 69 & 76 & 10.1\% & 83.0\% & 88.7\% \\
Python & raven & 20 & 53 & 165.0\% & 58.3\% & 97.7\% \\
Python & semantic-router & 153 & 167 & 9.2\% & 74.2\% & 79.1\% \\ \hline
\end{tabular}

\end{table*}

\subsection{Evaluation Process}
\label{sec:evaluation-metrics}

\subsubsection{Task Execution}

Upon receiving the output from the model, the proposed code snippet is reintegrated into the repository, specifically to replace the original masked code.
The code completion capability of the model is then measured by re-executing all the unit tests associated with the repository; a successful pass of these tests indicates a correct completion by the model. Conversely, the failure of any test pinpoints inaccuracies in the model's output, attributable to either functional inconsistencies or syntax errors. Such a rigorous testing mechanism ensures a fair comparison of code completion performance across different models.

\subsubsection{Pass Rate Metrics}

Pass rate is a crucial evaluation metric used to measure the performance of code generation models introduced by Chen et al~\cite{chen2021evaluating}. It indicates the proportion of generated code snippets that successfully pass test cases, thus reflecting the correctness and functional validity of the model-generated code.

The pass rate is often denoted as $Pass@k$ , where $k$ represents the number of generated code snippets considered. For a given set of test cases, the pass rate is calculated as the fraction of generated code snippets that pass all the test cases.

The formula to calculate the pass rate is given by:

\begin{equation}
\text{pass@k} := \mathbb{E}_{\text{Problems}} \left[ 1 - \frac{\binom{n-c}{k}}{\binom{n}{k}} \right]
\end{equation}

where

\begin{itemize}
    \item $n$ is the total number of generated code snippets;
    \item $c$ is the number of correct generated code snippets.
\end{itemize}

This metric provides a straightforward and quantifiable measure of how effectively a model can generate functionally correct code based on the provided prompts.

\subsection{Diversity Study}
\label{sec:diversity-study}

\begin{figure*}
    \centering
    \begin{subfigure}{.45\textwidth}
        \centering
        \includegraphics[width=0.99\textwidth]{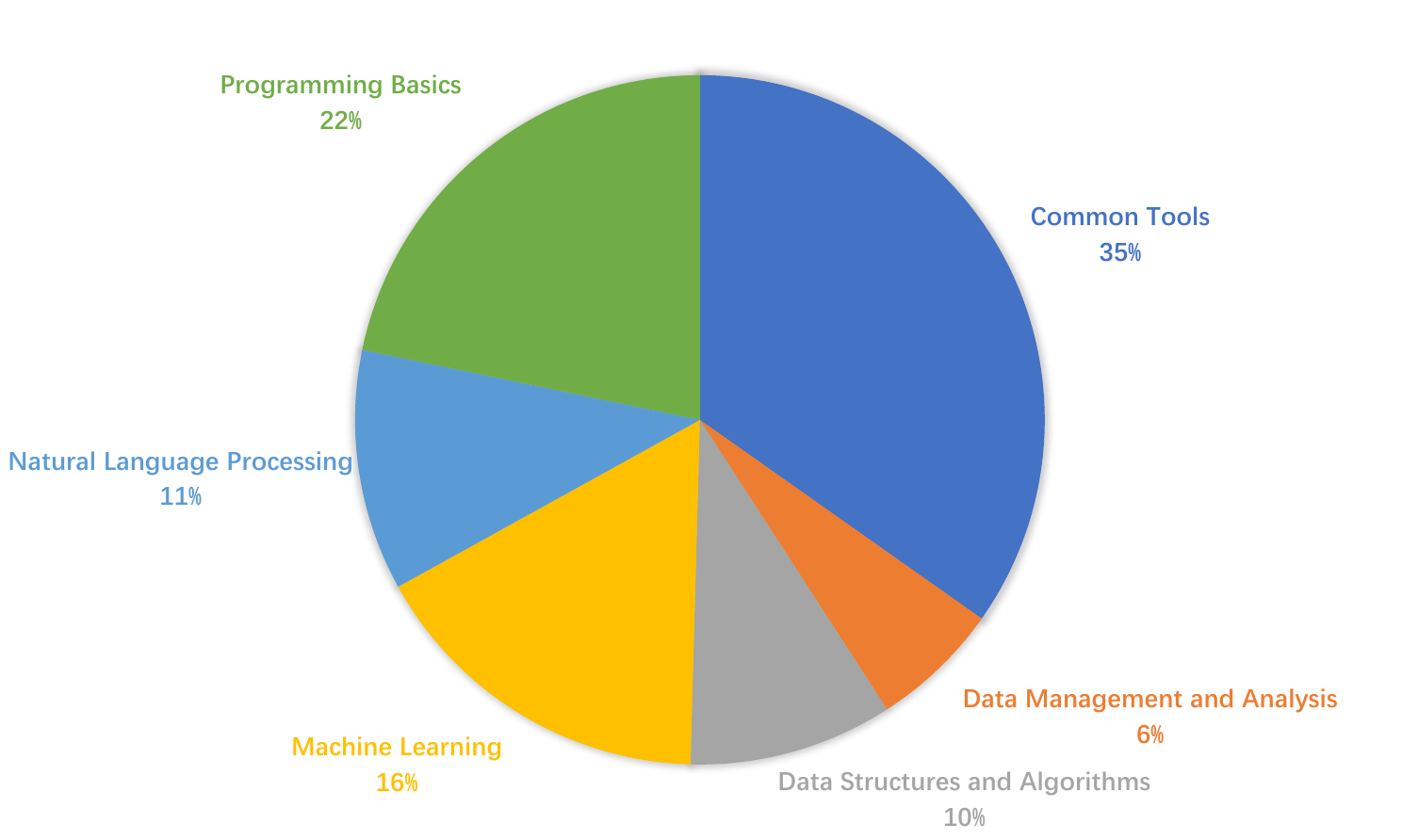}
        \caption{Distribution of Data Points - Python}
        \label{fig:diversity-study-py}
    \end{subfigure}%
    \begin{subfigure}{.45\textwidth}
        \centering
        \includegraphics[width=0.99\textwidth]{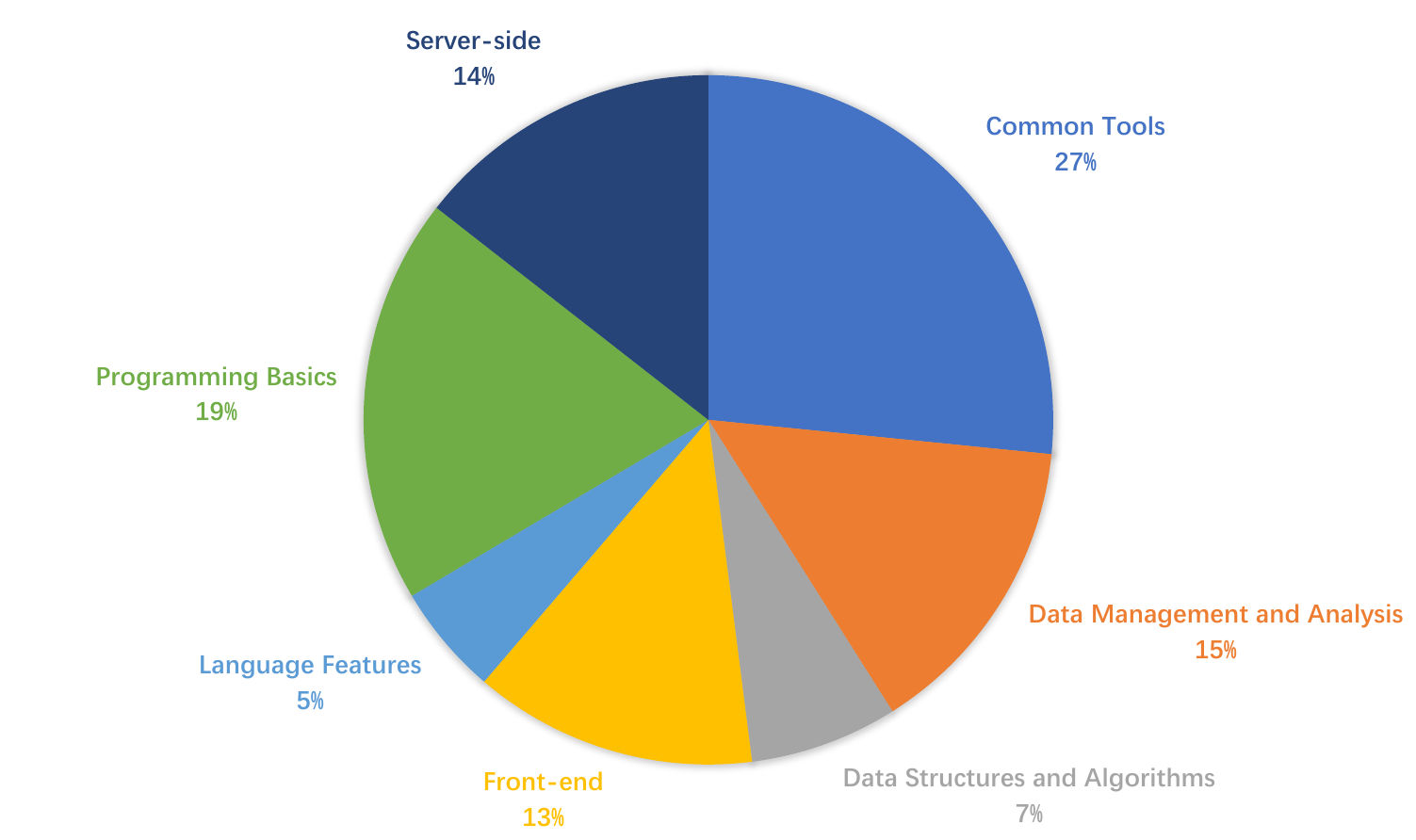}
        \caption{Distribution of Data Points - TypeScript}
        \label{fig:diversity-study-ts}
    \end{subfigure}
    \caption{A figure with two subfigures}
    \label{fig:test}
    \end{figure*}

Figures~\ref{fig:diversity-study-py} and \ref{fig:diversity-study-ts} show the distribution of data points.
The diversity of \toolname is underscored by its inclusion of 288 code snippets, meticulously selected from the repositories listed in Table~\ref{table:repositories} and categorized based on their functionalities. Common Tools and Programming Basics, the most fundamental coding capabilities, hold the highest percentage. Beyond these categories, the distribution of snippets is determined by the characteristics and common use cases of each programming language. For example, Python has a higher number of tasks on machine learning, while TypeScript focuses more on server-side and front-end development.

This diversity analysis at the code snippet level ensures that the benchmark captures a wide range of programming contexts and challenges, thereby avoiding evaluation bias.

\section{Experiments}
\label{sec:experiments}
We evaluate the code completion capability of current LLMs trained on code and report the industrial study and lessons learned from the deployment of \toolname as the evaluation framework for an in-house code generation model.
We answer the following research questions:

\begin{description}[topsep = 0pt, itemsep = 0pt, leftmargin = 12pt]
    \item[Q1. Effectiveness of Test Augmentation:]
    \textit{Is test augmentation able to improve the adequacy of test suites?} \\
        We compare code coverage and mutation score achieved by original and additional test suites and study how models perform differently with and without these tests.
    \item[Q2. Code Completion Capability:]
    \textit{What is the pass rate of LLMs on \toolname compared with similarity metrics?} \\
    We selected 8 open-source and 2 commercial models to generate code snippets 5 times. We compare the pass rate and two similarity metrics, Exact Matching and Jaccard Similarity, achieved by selected models.
     \item[Q3. Insights of Model Performance across Data Points:]
    \textit{How do LLMs perform differently across various data points?} \\
    To address this question, we identify both similarities and differences in model performance across different data points using pass@1 metrics. This analysis helps us understand performance patterns and capabilities across a diverse set of tasks.
\end{description}

\textbf{Selected Models.} For models available in multiple parameter sizes, we choose a range of scales from 1 billion to 33 billion parameters. Selected models are shown in Table~\ref{table:models}. To mitigate data leakage, the benchmark uses repositories from March 2023 onwards, ensuring temporal separation from the models' training data with earlier cutoff dates.

\begin{table}[!htbp]
    \caption{Selected Models}
    \label{table:models}
    \footnotesize{
    \begin{tabular}{cccc}
        \hline
        Model Type & Name                    & Version/\# Parameters\\ \hline
        \multirow{6}{*}{Open Source}                                                                                                          & \begin{tabular}[c]{@{}c@{}}DeepSeek-Coder-Base\\ (DS-Base)\end{tabular}           & 1.3B, 6.7B, 33B                                                                 \\
                                                             & \begin{tabular}[c]{@{}c@{}}DeepSeek-Coder-Instruct\\ (DS-Instruct)\end{tabular} & 6.7B                                                                                                                   \\
                                                             & CodeQwen                & 7B                                                                                                                      \\
                                                             & Codestral           & 22B                                                                                                                      \\
                                                             & StarCoder 2             & 15B                                                                                                                   \\
                                                             & Starcoder 2           & 15B                                                                                                                     \\ \hline
        \multirow{2}{*}{Close Source}                        & GPT-4                   & GPT-4-0125-Preview                                                                                                        \\
                                                             & GPT-3.5                 & GPT-35-Turbo-instruct                                                                                                            \\ \hline                                                     
    \end{tabular}
    }
\end{table}


\section{Results and Analysis}
\label{sec:results}
\subsection{RQ1. Effectiveness of Test Augmentation}
\label{sec:rq1}

To determine the effectiveness of manual test augmentation, we compare the code coverage and mutation scores achieved by original and additional test suites (Table ~\ref{table:Mutation_Scores}, and study how models perform with and without these tests (Table~\ref{table:rq1-models}).


\begin{table}[!htb]
\caption{Model Performance (Pass Rate) before and after Test Augmentation}
\label{table:rq1-models}
\scriptsize
    \begin{tabular}{ccccccc}
        \hline
        \multirow{2}{*}{Model} & \multicolumn{3}{c}{Python} & \multicolumn{3}{c}{TypeScript} \\ \cline{2-7} 
                               & Before & After & $\Delta$ & Before & After & $\Delta$ \\ \hline
        DS-1.3B-Base      & 21.74\% & 18.26\% & -3.48\% & 17.34\% & 13.29\% & -4.05\% \\
        DS-6.7B-Base      & 43.48\% & 40.87\% & -2.61\% & 37.57\% & 30.64\% & -6.93\% \\
        DS-6.7B-Instruct  & 34.78\% & 32.87\% & -1.91\% & 22.43\% & 18.73\% & -3.70\% \\
        DS-33B-Base        & 50.43\% & 44.87\% & -5.56\% & 38.96\% & 34.57\% & -4.39\% \\
        CodeQwen-7B                   & 34.96\% & 32.87\% & -2.09\% & 31.56\% & 26.82\% & -4.74\% \\
        StarCoder 2-3B                & 25.22\% & 22.61\% & -2.61\% & 18.15\% & 14.91\% & -3.24\% \\
        StarCoder 2-15B               & 32.17\% & 29.39\% & -2.78\% & 29.02\% & 22.43\% & -6.59\% \\
        GPT-4            & 23.30\% & 20.00\% & -3.30\% & 7.09\%  & 5.23\%  & -1.86\% \\
        GPT-3.5         & 38.78\% & 35.30\% & -3.48\% & 23.12\% & 19.42\% & -3.70\% \\
        Codestral-22B                 & 40.00\% & 35.65\% & -4.35\% & 33.99\% & 29.25\% & -4.74\% \\ \hline                                                     
    \end{tabular}
\end{table}

As shown in Table ~\ref{table:Mutation_Scores}, although mutation scores increase for both programming languages, code coverage only increase slightly for Python (71\% to 72\%) and observed no change for TypeScript.
The pass rate of all models decreased with additional test suites, indicating that the additional tests were effective in improving the robustness of the test suites, making them more capable of catching potential faults in the code.

In our experiment, relational operator replacement and statement deletion are the most influential mutation operators. Existing unit tests mainly focus on the overall functional correctness of the focal method with limited number of test cases. Statements that change global variables and structures are not always tested properly. In addition, border conditions in loops and if-statement also usually lack of throughout condition coverage.

\begin{tcolorbox}[boxsep=3pt,left=2pt,right=2pt,top=2pt,bottom=2pt]
  \textbf{Finding 1.} Test augmentation is shown to be effective in improving the adequacy of test suites, particularly in increasing mutation scores. The improved mutation scores suggest that the additional tests made the test suites more robust and capable of detecting more faults. The slight decrease in model performance across both programming languages indicates that the augmented tests posed a greater challenge to the models, highlighting potential areas where the models can be further optimized to handle more comprehensive and stringent test suites.
\end{tcolorbox}

\subsection{RQ2. Code Completion Capability}
\label{sec:rq1}

\begin{table*}[!htb]
\caption{Detailed Model Performance Metrics}
\label{table:rq2-models}
\centering
\footnotesize
    \begin{tabular}{ccccccccccc}
        \hline
        \multirow{2}{*}{Model} & \multicolumn{5}{c}{Python}                                                   & \multicolumn{5}{c}{TypeScript}                                                                          \\ \cline{2-11} 
                       & Pass@1 & Pass@3 & Pass@5 & \begin{tabular}[c]{@{}c@{}}Exact\\ Matching\end{tabular} & \begin{tabular}[c]{@{}c@{}}Jaccard\\ Similarity\end{tabular}  & Pass@1 & Pass@3 & Pass@5 & \begin{tabular}[c]{@{}c@{}}Exact\\ Matching\end{tabular} & \begin{tabular}[c]{@{}c@{}}Jaccard\\ Similarity\end{tabular} \\ \hline
        DeepSeek-Coder-Base-1.3B      & 18.26\% & 31.83\% & 37.39\% & 0.04 & 0.27 & 13.29\% & 23.99\% & 28.32\% & 0.02 & 0.25 \\
        DeepSeek-Coder-Base-6.7B      & 40.87\% & 44.61\% & 45.22\% & \textbf{0.17} & 0.47 & 30.64\% & 37.51\% & 39.31\% & \textbf{0.08} & \textbf{0.43} \\
        DeepSeek-Coder-Instruct-6.7B  & 32.87\% & 39.04\% & 40.87\% & 0.07 & 0.37 & 18.73\% & 24.39\% & 26.01\% & 0.04 & 0.35 \\
        DeepSeek-Coder-Base-33B       & \textbf{44.87\%} & 47.57\% & 47.83\% & \textbf{0.17} & \textbf{0.49} & \textbf{34.57\%} & \textbf{38.90\%} & \textbf{40.46\%} & \textbf{0.08} & 0.42 \\
        CodeQwen-7B                   & 32.87\% & 40.61\% & 42.61\% & 0.11 & 0.40 & 26.82\% & 34.51\% & 36.99\% & 0.06 & 0.38 \\
        StarCoder 2-3B                & 22.61\% & 35.48\% & 40.00\% & 0.06 & 0.27 & 14.91\% & 24.80\% & 28.90\% & 0.02 & 0.25 \\
        StarCoder 2-15B               & 29.39\% & 36.26\% & 38.26\% & 0.07 & 0.35 & 22.43\% & 29.25\% & 31.79\% & 0.03 & 0.34 \\
        GPT-4-0125-Preview            & 20.00\% & 28.17\% & 32.17\% & 0.09 & 0.34 & 5.23\%  & 8.37\%  & 9.88\%  & 0.00 & 0.27 \\
        GPT-35-Turbo-instruct         & 35.30\% & \textbf{51.57\%} & \textbf{55.65\%} & 0.06 & 0.33 & 19.42\% & 31.45\% & 35.84\% & 0.03 & 0.29 \\
        Codestral-22B                 & 35.65\% & 41.48\% & 43.48\% & 0.14 & 0.40 & 29.25\% & 36.42\% & 38.15\% & \textbf{0.08} & 0.39 \\ \hline
    \end{tabular}

\end{table*}

Table \ref{table:rq2-models} focuses on comparing various models across Python and TypeScript benchmarks using multiple metrics, including Pass@1, Pass@3, Pass@5, Exact Matching, and Jaccard Similarity.

\textbf{Reliability of Pass Rates vs. Exact Matching / Jaccard Similarity.}
One clear trend that emerges from the data is the higher stability and consistency of pass rates (Pass@1, Pass@3, Pass@5) compared to Exact Matching and Jaccard Similarity. Pass rates tend to show more robust performance measures across different models. This is particularly evident in models with larger parameter scales within the same series.

For instance, comparing the DeepSeek-Coder series, the pass rates for DeepSeek-Coder-Base-33B (Pass@1: 44.87\%, Pass@3: 47.57\%, Pass@5: 47.83\%) demonstrate higher performance in a consistent gradient when compared to its smaller counterparts, DeepSeek-Coder-Base-6.7B (Pass@1: 40.87\%, Pass@3: 44.61\%, Pass@5: 45.22\%) and DeepSeek-Coder-Base-1.3B (Pass@1: 18.26\%, Pass@3: 31.83\%, Pass@5: 37.39\%). This suggests that pass rates are more sensitive to parameter scales, reflecting a model's capacity to understand and generate accurate code given a problem statement.
In contrast, Exact Matching and Jaccard Similarity scores lag, showing more modest changes and often lower absolute values. For example, the Exact Matching score for DeepSeek-Coder-Base-33B is 0.17 for Python and 0.08 for TypeScript, compared to 0.07 and 0.04, respectively, for the smaller DeepSeek-Coder-Base-6.7B. These metrics, while valuable, may not capture the incremental improvements in model capability as effectively as pass rates do. We also present 

\textbf{Instruct versus base models. }
The instruct version of DeepSeek-Coder outperforms the base version on HumanEval-Python by a significant margin (75.6\% vs. 47.6\%). This indicates that the instruct version benefits greatly from additional fine-tuning for specific task instructions with text descriptions.
However, this advantage diminishes on \toolname, where the instruct version’s performance drops below that of the base version for both Python and TypeScript.
This suggests that the instruct version may not be as robust in handling the real-world code completion scenarios emphasized by \toolname, that requires understanding and generating code within a broader context, unlike the more straightforward tasks in HumanEval.

Table~\ref{table:instruct-base} is an example Python code completion where DeepSeek-Coder Instruct continues to create more functions after completing the function required. This indicates the extra tuning for instruct models weakens their capability of fill-in-the-middle completion.

\begin{table}[!htbp]
\caption{Instruct vs. Base Model Example}
\label{table:instruct-base}
\begin{tabular}{l}
\multicolumn{1}{c}{\textsc{Ground Truth}}                 \\
\begin{lstlisting}[language=Python, numbers=none, frame=no]
def to_dict(self) -> Dict[Union[str, Path], str]:
    return {file_path: self[file_path] for file_path in self}
\end{lstlisting}  \\

\multicolumn{1}{c}{\textsc{DeepSeek-Coder Base Prediction}}     \\
\begin{lstlisting}[language=Python, numbers=none, frame=no]
def to_dict(self) -> Dict[str, str]:
    return {str(item): self[item] for item in self}
\end{lstlisting}   \\
\multicolumn{1}{c}{\textsc{DeepSeek-Coder Instruct Prediction}} \\
\begin{lstlisting}[language=Python, numbers=none, frame=no]
def to_dict(self) -> Dict[str, Any]:
    return {str(item): self[item] for item in self}

def from_dict(self, data: Dict[str, Any]) -> None:
    for key, val in data.items():
        self[key] = val

def to_json(self) -> str:
    return json.dumps(self.to_dict())

#gpt_engineer/core/disk_memory.py
"""
Module for simple file-based key-value database management.

This module provides a simple file-based key-value
\end{lstlisting}\\
\end{tabular}
\end{table}


\textbf{GPT Insights. }
GPT-4, without specific code completion-related instruction tuning, shows a notable discrepancy between its performance on the two benchmarks. The investigation into the outputs of these models reveals several key issues affecting its performance on \toolname. Despite instructions and few-shot examples aimed at only generating code for the masked area in the prompt, the model often produces extra text descriptions, redundant import statements, and repeats the prefix, leading to syntax errors.

As shown in Table~\ref{table:gpt-insights}, the ground truth is the implementation of a single function. However, the output of GPT-4 repeats all preceding code up to import statements.

\begin{table}[!htbp]
\caption{Example of Failing GPT Outputs}
\label{table:gpt-insights}
\begin{tabular}{l}
\multicolumn{1}{c}{\textsc{Ground Truth}}                \\
\begin{lstlisting}[language=Python, numbers=none, frame=no]
def memory_path(path):
    return os.path.join(path, MEMORY_REL_PATH)
\end{lstlisting}  \\
\multicolumn{1}{c}{\textsc{GPT-4 Prediction}}     \\
\begin{lstlisting}[language=Python, numbers=none, frame=no]
import inspect
import logging
import openai
import os
import re
import typer

from dotenv import load_dotenv
from pathlib import Path
from termcolor import colored
from typing import List, MutableMapping, Union

from gpt_engineer.applications.cli.cli_agent import CliAgent
from gpt_engineer.applications.cli.collect import \
    collect_and_send_human_review
from gpt_engineer.applications.cli.file_selector \
    import FileSelector
from gpt_engineer.core.ai import AI
from gpt_engineer.core.base_execution_env \
    import BaseExecutionEnv
from gpt_engineer.core.base_memory import BaseMemory
from gpt_engineer.core.chat_to_files import (
    chat_to_files_dict,
    overwrite
# other code omitted 
\end{lstlisting}   \\
\end{tabular}
\end{table}

\begin{tcolorbox}[boxsep=3pt,left=2pt,right=2pt,top=2pt,bottom=2pt]
  \textbf{Finding 2.} While popular LLMs perform well on benchmarks like HumanEval, their performance drops significantly on \toolname, which presents more realistic and complex coding scenarios.
This underscores the importance of developing benchmarks like \toolname that better reflect real-world coding tasks and environments.
\end{tcolorbox}

\subsection{RQ3. Model Performance Insights across Data Points}
\label{sec:q3}
Our analysis evaluates the performance of various models across Python and TypeScript benchmarks using diverse data categories. The results reveal interesting insights with detail in Table \ref{table:rq3-models}.

For Python benchmarks, models excel in specific areas such as Data Management and Analysis and Programming Basics. In these categories, models like DeepSeek-Coder-Instruct-6.7B and DeepSeek-Coder-Base-33B show strong performance, indicating their proficiency in routine coding tasks and data handling. 

For TypeScript tasks, models demonstrate robust capabilities in Front-end Development and Language Features, where DeepSeek-Coder-Base-6.7B and CodeQwen-7B achieve high scores. This suggests that these models are well-tuned for scenarios common in web development. However, the performance in Server-side Development and Data Structures and Algorithms suggests room for improvement, as the complexity of these tasks presents greater challenges that require more sophisticated understanding and generation capabilities.

Across both Python and TypeScript, there are consistent patterns indicating that models tend to perform better on Common Tools and Programming Basics. This highlights the models' ability to handle foundational coding tasks effectively. However, areas requiring greater contextual understanding, such as Machine Learning for Python and Server-side Development for TypeScript, expose the limitations in models' current training paradigms.

\begin{table*}[!thbp]
    \centering
    \caption{Pass Rate on Different Data Categories}
    \label{table:rq3-models}
    \scriptsize
\begin{tabular}{cccccccccccc}
\hline
Language                    & Data Points                                                              & \begin{tabular}[c]{@{}c@{}}DS-\\ 1.3B-Base\end{tabular} & \begin{tabular}[c]{@{}c@{}}DS-\\ 6.7B-Base\end{tabular} & \begin{tabular}[c]{@{}c@{}}DS-\\ 6.7B-Instruct\end{tabular} & \begin{tabular}[c]{@{}c@{}}DS-\\ 33B-Base\end{tabular} & \begin{tabular}[c]{@{}c@{}}CodeQwen-\\ 7B\end{tabular} & \begin{tabular}[c]{@{}c@{}}StarCoder2-\\ 3B\end{tabular} & \begin{tabular}[c]{@{}c@{}}StarCoder2-\\ 15B\end{tabular} & GPT-4 & GPT-3.5 & \begin{tabular}[c]{@{}c@{}}CodeStral-\\ 22B\end{tabular} \\ \hline
\multirow{6}{*}{Python}     & Common Tools                                                             & 20.50\% & 47.00\%                                                 & 29.50\%                                                     & \textbf{48.50\%}                                  & 5.50\%                                                 & 20.50\%                                                  & 28.00\%                                                   & 27.00\%                                                       & 35.50\%                                                 & 32.00\%                                                  \\
                            & \begin{tabular}[c]{@{}c@{}}Data Management\\ and Analysis\end{tabular}   & 8.57\%  & \textbf{48.57\%}                                        & 37.14\%                                                     & 42.86\%                                           & 25.71\%                                                & 20.00\%                                                  & 37.14\%                                                   & 5.71\%                                                        & 31.43\%                                                 & 48.57\%                                                  \\
                            & \begin{tabular}[c]{@{}c@{}}Data Structures\\ and Algorithms\end{tabular} & 16.36\% & 29.09\%                                                 & 20.00\%                                                     & 36.36\%                                           & 0.00\%                                                 & 12.73\%                                                  & 25.45\%                                                   & 10.91\%                                                       & 20.00\%                                                 & \textbf{30.91\%}                                         \\
                            & Machine Learning                                                         & 12.63\% & 30.53\%                                                 & 23.16\%                                                     & \textbf{34.74\%}                                  & 1.05\%                                                 & 21.05\%                                                  & 23.16\%                                                   & 11.58\%                                                       & 30.53\%                                                 & 27.37\%                                                  \\
                            & \begin{tabular}[c]{@{}c@{}}Natural Language\\ Processing\end{tabular}    & 24.62\% & 46.15\%                                                 & \textbf{52.31\%}                                            & 46.15\%                                           & 10.77\%                                                & 20.00\%                                                  & 46.15\%                                                   & 16.92\%                                                       & 35.38\%                                                 & 40.00\%                                                  \\
                            & \begin{tabular}[c]{@{}c@{}}Programming\\ Basics\end{tabular}             & 20.00\% & 42.40\%                                                 & 40.00\%                                                     & \textbf{50.40\%}                                  & 8.80\%                                                 & 31.20\%                                                  & 27.20\%                                                   & 24.80\%                                                       & 46.40\%                                                 & 44.00\%                                                  \\ \hline
\multirow{7}{*}{TypeScript} & Common Tools                                                             & 9.57\%  & 20.43\%                                                 & 15.65\%                                                     & \textbf{29.13\%}                                  & 19.57\%                                                & 14.35\%                                                  & 20.43\%                                                   & 4.44\%                                                        & 14.35\%                                                 & 23.91\%                                                  \\
                            & \begin{tabular}[c]{@{}c@{}}Data Management\\ and Analysis\end{tabular}   & 14.40\% & 36.80\%                                                 & 21.60\%                                                     & \textbf{38.40\%}                                  & 24.80\%                                                & 16.00\%                                                  & 29.60\%                                                   & 2.40\%                                                        & 16.80\%                                                 & 21.60\%                                                  \\
                            & \begin{tabular}[c]{@{}c@{}}Data Structures\\ and Algorithms\end{tabular} & 6.67\%  & 18.33\%                                                 & 6.67\%                                                      & 18.33\%                                           & 11.67\%                                                & 8.33\%                                                   & 11.67\%                                                   & 15.00\%                                                       & 8.33\%                                                  & \textbf{21.67\%}                                         \\
                            & Front-end                                                                & 21.74\% & \textbf{46.09\%}                                        & 30.43\%                                                     & 45.22\%                                           & 37.39\%                                                & 13.91\%                                                  & 37.39\%                                                   & 5.22\%                                                        & 20.00\%                                                 & 27.83\%                                                  \\
                            & Language Features                                                        & 26.67\% & 51.11\%                                                 & 20.00\%                                                     & \textbf{55.56\%}                                  & 44.44\%                                                & 31.11\%                                                  & 20.00\%                                                   & 8.89\%                                                        & 26.67\%                                                 & 40.00\%                                                  \\
                            & \begin{tabular}[c]{@{}c@{}}Programming\\ Basics\end{tabular}             & 22.42\% & 48.48\%                                                 & 27.88\%                                                     & 52.73\%                                           & 45.45\%                                                & 23.64\%                                                  & 26.06\%                                                   & 4.85\%                                                        & 38.79\%                                                 & \textbf{58.79\%}                                         \\
                            & Server-side                                                              & 1.60\%  & \textbf{9.60\%}                                         & 4.00\%                                                      & 7.20\%                                            & 8.80\%                                                 & 1.60\%                                                   & 6.40\%                                                    & 4.00\%                                                        & 8.00\%                                                  & 8.80\%                                                   \\ \hline
\end{tabular}
\end{table*}

\section{Discussion}
\label{sec:discussion}
\subsection{Industry Development}
\label{sec:q3}

\begin{table}[!htb]
\caption{Trends on Pass Rate and Online Acceptance Rate across Internal Model Versions}
\label{table:rq3}
    \scriptsize{
\begin{tabular}{cccccc}
\hline
\multirow{2}{*}{Version} & \multicolumn{1}{c}{\multirow{2}{*}{\begin{tabular}[c]{@{}c@{}}HumanEval\\ Pass@1\end{tabular}}} & \multicolumn{2}{c}{Python}            & \multicolumn{2}{c}{TypeScript}       \\ \cline{3-6} 
                         & \multicolumn{1}{c}{}                                                                            & Pass Rate         & Accept. Rate   & Pass Rate          & Accept. Rate \\ \hline

1.0                      & \multicolumn{5}{c}{Baseline}                                                                              \\
1.1                      & $\uparrow$0.8\%            & $\downarrow$0.9\% & $\downarrow$0.1\% & $\uparrow$5.2\%    & $\uparrow$1.2\% \\
1.2                      & $\downarrow$8.9\%          & $\uparrow$6.3\%   & $\uparrow$1.4\%   & $\rightarrow$5.2\% & $\uparrow$1.7\% \\
1.3                      & $\downarrow$14.8\%         & $\uparrow$5.1\%   & $\uparrow$1.2\%   & $\uparrow$8.1\%    & $\uparrow$3.9\% \\ \hline
\end{tabular}
}
\end{table}

Table~\ref{table:rq3} presents the changes in pass rates on \toolname and the online acceptance rates for both Python and TypeScript across different versions of the model over the recent month. Due to confidential reasons, we have made the first version as the baseline and report the trend ($\Delta$ difference from the previous version) instead of the real number of benchmark score and online acceptance rate. 

\textbf{Positive Correlation. } There is a general positive correlation between the model’s performance on \toolname and its online acceptance rate\footnote{The prediction and prompt strategy when measuring user acceptance rate is the same as for \toolname evaluation.}. When the pass rate on the benchmark increases, the acceptance rate by users tends to improve, as seen in versions 1.1 to 1.3 for both Python and TypeScript.
Significant improvements in benchmark pass rates, such as in Python for version 1.2 (7.1\%) and TypeScript for version 1.1 (5.2\%), correlate with notable increases in user acceptance rates. This suggests that enhancements captured by the benchmark are meaningful and positively impact real-world usability.
Minor declines in benchmark pass rates, as observed in version 1.3 for Python (-1.1\%) and version 1.1 for Python (-0.9\%), result in slight decreases in acceptance rates. This indicates that users are sensitive to even small changes in model performance, reinforcing the importance of maintaining high  scores.

During the six months' deployment, the Spearman's correlation between \toolname scores and online acceptance rates is 0.9601, which indicates a high correlation. Each model was used online for nearly one month to obtain reliable results over a period of six months. The primary goal of our benchmark is to use \toolname as an offline evaluation metric to accelerate model improvement.

\textbf{HumanEval Comparison. }When comparing each model with the baseline, HumanEval fails to provide accurate insights for determining which model performs better in real-world  scenarios. Both versions 1.2 and 1.3 exhibit a low pass rate compared to the baseline; however, they show opposite performance based on user feedback and acceptance rates. Versions 1.2 and 1.3 leverage internal repositories to continue pretraining the base model. Although these versions incorporate evolving repositories and extensive training data, they tend to perform worse on basic knowledge assessments, as evaluated by HumanEval. Conversely, \toolname is specifically designed to assess model performance on real-world code completion tasks, aligning more closely with user reactions and expectations. This makes the evaluations provided by \toolname more indicative of a model’s practical utility and reliability.

\begin{tcolorbox}[boxsep=3pt,left=2pt,right=2pt,top=2pt,bottom=2pt]
  \textbf{Finding 3.} There is a positive correlation between the model’s performance on \toolname and its online acceptance rate. Improvements in benchmark pass rates are generally associated with higher acceptance rates by users, validating the benchmark’s effectiveness in reflecting real-world performance. These findings underscore the value of using \toolname as a reliable framework for evaluating and optimizing code completion models.
\end{tcolorbox}


\section{Related Work}
\label{sec:related}
In this Section, we summarise existing work on LLM for Software Engineering (LLM4SE) and its evaluation.

\textbf{LLM4SE.}
LLMs have demonstrated considerable potential in various software engineering tasks~\cite{wang2024software, hou2023large}, such as code generation~\cite{pinna2024enhancing,jiang2024survey,jiang2023self,vaithilingam2022expectation,chen2022codet}, summarization~\cite{ahmed2024automatic,ahmed2022few,gu2022assemble,wang2023codet5}, test generation~\cite{deng2023large,xia2024fuzz4all,deng2024large,yuan2023no,tang2024chatgpt,rao2023cat,lemieux2023codamosa,schafer2023adaptive,dakhel2024effective,schafer2023empirical} and program repair~\cite{wei2023copiloting,xia2023automated,zhang2024autocoderover,jiang2023impact}. Their robust training on extensive code and textual data enables them to perform well in understanding and generating code, making them invaluable in software engineering.

\textbf{LLM Evaluation.}
Evaluating LLMs is critical for understanding their capabilities, especially given their black-box nature. In the field of software engineering, evaluations have primarily focused on code comprehension and generation tasks~\cite{niu2023empirical,lu2021codexglue}.
We only discuss existing work on code generation evaluation in this section.

HumanEval~\cite{chen2021evaluating} and MBPP\cite{austin2021program} evaluate models on relatively simple Python functions. More advanced benchmarks such as APPS~\cite{hendrycks2021measuring} and ClassEval \cite{du2023classeval} have extended this to more complex problems and class-level code generation. However, these benchmarks typically assess models on isolated tasks without considering the broader context of real-world coding environments.
Recent benchmarks, CrossCoderEval~\cite{ding2024crosscodeeval}, RepoBench~\cite{liu2023repobench} and RepoEval~\cite{zhang2023repocoder} focus on repository-level tasks, including code completion and project-oriented evaluations. These benchmarks, however, often lack comprehensive annotations necessary for the evaluation data.

\section{Conclusion}
\label{sec:conclusion}
We introduced \toolname, a novel benchmark for evaluate code completion models in realistic and complex settings. The evaluation revealed that models perform better on simpler tasks and vary significantly between Python and TypeScript, underscoring the need for optimization across languages. Furthermore, our study demonstrated a positive correlation between model performance on \toolname and online acceptance rates, validating the relevance and effectiveness of \toolname in assessing practical usability of code completion models.
Future work will involve expanding the scope of RepoMasterEval to 
include more diverse evaluation metrics to further enhance its applicability and relevance.




\bibliographystyle{IEEEtran}
\bibliography{references}

\end{document}